\documentclass[secnumarabic,amssymb,aps]{revtex4-2}
\usepackage{epstopdf,float}
\usepackage[utf8]{inputenc}
\usepackage{amsmath,amssymb,amsthm,amsfonts,mathrsfs,bm,verbatim}
\usepackage{graphicx,subfigure,diagbox}
\usepackage[section]{placeins}
 \usepackage{booktabs} 
\usepackage{hyperref}
\hypersetup{
	colorlinks=true,
	linkcolor=blue,
	citecolor=blue 
}
\usepackage{ulem}
\newcommand{\bea}{\begin{eqnarray}}	\newcommand{\eea}{\end{eqnarray}}

\allowdisplaybreaks[4]

\UseRawInputEncoding

\begin{document}

	\title{Scalar quasinormal modes, Lyapunov exponents and radii of null geodesics of rotating regular black holes}
	
	\author{Yating Peng $^{1}$}
	\author{Jia-Hui Huang$^{1,2}$}
	\email{huangjh@m.scnu.edu.cn}
	
	\affiliation{
		$^1$ Key Laboratory of Atomic and Subatomic Structure and Quantum Control (Ministry of Education), Guangdong Basic Research Center of Excellence for Structure and Fundamental Interactions of Matter, School of Physics, South China Normal University, Guangzhou 510006, China\\
		$^2$ Guangdong Provincial Key Laboratory of Quantum Engineering and Quantum Materials, Guangdong-Hong Kong Joint Laboratory of Quantum Matter, South China Normal University, Guangzhou 510006, China}
	
	\begin{abstract}
The quasinormal modes of massless scalar field in rotating Bardeen and Hayward regular black holes are studied. The fundamental quasinormal modes and $n=1$ first overtone quasinormal modes are calculated with two numerical methods, i.e. WKB method and matrix method. Impacts of model parameters on the quasinormal modes are also discussed. It is found that the quasinormal modes in rotating Hayward black hole have just percent-level increase compared with that in the Kerr black hole case due to the deviation parameter $g$, while there is ten-percent-level increase for the quasinormal modes in the rotating Bardeen black hole due to $g_*$. It is also found that the monotonicity of fundamental and $n=1$ quasinormal modes in the rotating Bardeen black hole is different. The corotating and counterrotating Lyapunov exponents of the equatorial null circular geodesics in the two rotating black holes are also calculated. Then, the connection between the imaginary parts and real parts of the eikonal quasinormal modes, particularly the first overtone ones, and the properties of the null geodesics in the two rotating regular black holes are explicitly verified.  

	\end{abstract}

	\maketitle

	\section{Introduction}\label{section1}

Quasinormal modes (QNMs) play important roles in the study of dissipative systems. QNMs are complex-frequency vibrations with the real part indicating frequency and the imaginary part representing decay rate \cite{Vishveshwara:1970zz,Berti:2009kk,Kokkotas:1999bd,Konoplya:2011qq}. In general relativity, QNMs describe how black holes respond to perturbations, and their spectra depend on black hole parameters like mass and spin.This makes them a crucial tool for probing black hole characteristics. Regge and Wheeler \cite{Regge:1957td} analyzed Schwarzschild black hole perturbations, and Teukolsky \cite{Teukolsky:1972my} separated Kerr black hole perturbations using the Newman-Penrose formalism. In gravitational wave astronomy, QNMs help measure black hole parameters and test the no-hair theorem \cite{Capano:2021etf}. In gauge/gravity duality \cite{Kokkotas:1999bd}, QNMs relate to transport coefficients in strongly coupled quantum field theories, e.g. aiding research on quark-gluon plasma.
	
Another important aspect of black hole physics is the study of their shadows \cite{Cunha:2018acu,Dokuchaev:2019jqq,Perlick:2021aok,Wang:2022kvg}, which are the apparent images of the photon region around a black hole. The shadow's shape and size depend on the black hole's parameters and its spacetime geometry. Recent work has established a connection between the real part of eikonal $(l\gg 1)$ QNMs and the black hole shadow radius \cite{Jusufi:2019ltj,Yang:2021zqy}. 
This connection has been investigated for many black hole models \cite{Cuadros-Melgar:2020kqn,Guo:2020nci,Jusufi:2020mmy,Hendi:2020knv,Jusufi:2020agr,Cai:2020kue,Jusufi:2020odz,Jusufi:2020wmp,Mondal:2020pop,Saurabh:2020zqg,Jafarzade:2020ova,Ghasemi-Nodehi:2020oiz,Cai:2020igv,Campos:2021sff,Cai:2021ele,Anacleto:2021qoe,Wu:2021pgf,Liu:2022plm,Konoplya:2022hll,Yu:2022yyv,Lambiase:2023hng,Yan:2023pxj,Das:2023ess,Konoplya:2023moy,Bolokhov:2023dxq,Gogoi:2024vcx,Giataganas:2024hil}
Particularly, this connection has been explored in detail for the Kerr black hole \cite{Jusufi:2020dhz,Yang:2021zqy}, Kerr-Sen black hole\cite{Wu:2021pgf} and charged AdS black hole in Gauss-Bonnet gravity \cite{Chaudhary:2021uuk}.

The connection between QNMs and Layapunov exponent\cite{Cardoso:2008bp,Johnson:2019ljv} was established by Ferrari and Mashhoon in 1984 and later was extended to stationary, spherically symmetric, and asymptotically flat metrics in \cite{Cardoso:2008bp}, where it was demonstrated that for a Schwarzschild black hole, the real component of QNMs in the eikonal limit is proportional to the angular velocity of the null circular orbit, while the imaginary component is associated with the Lyapunov exponent, which dictates the instability time scale of the orbit. The connection was also utilized to calculate the QNMs for both the noncommutative Schwarzschild black hole\cite{Giri:2022zhf} and dual stringy black hole\cite{Giri:2021kgc}.
     
Rotating regular black holes are important research objects in gravitational physics \cite{Bambi:2013ufa,:2022twv,Lan:2023cvz}, aiming to address the singularity problem in classical rotating black hole solutions such as the Kerr black hole. These regular models avoid the inner curvature singularities by incorporating modifications inspired by quantum gravity or other alternative approaches \cite{Bardeen1968,PhysRevLett.96.031103,Toshmatov:2014nya,Fan:2016hvf}. The interest in rotating regular black holes has been triggered by recent gravitational wave observations \cite{Middleton:2024ytu} and the imaging of black holes by the Event Horizon Telescope \cite{EventHorizonTelescope:2022xqj}. These developments provide a potential avenue to test gravitational theories and explore the nature of black holes beyond classical general relativity.

Recently, the connection between the quasinormal modes and shadow radius of the non-rotating Bardeen black hole in a cloud of strings has been discussed \cite{Vishvakarma:2023csw}.
In \cite{Pedrotti:2024znu}, the authors have extended a previous work on Kerr black hole \cite{Yang:2021zqy} to rotating regular black holes, and proposed a new formula to calculate the shadow radius with the real parts of the QNMs in rotating regular black holes. The connection between real parts of the fundamental QNMs and shadow radius in rotating Bardeen and Hayward black holes has been checked. 

In this work, we will consider the fundamental and first overtone QNMs in rotating Bardeen and Hayward black holes in detail. The methods used for calculating the QNMs are 3-order WKB method \cite{Iyer:1986np,Iyer:1986nq,Seidel:1989bp,Yang:2022xxh} and the matrix method \cite{Lin:2016sch,Lin:2017oag,Lin:2019mmf,Huang:2025wom}. We will calculate the Lyapunov exponents of the null geodesics and shadow radius of the black holes, and check the connection between the QNMs and the properties of the null geodesics of the two rotating regular black holes.

    The rest of this paper is then organized as follows: In Section 2, we introduce the metrics of rotating Hayward and Bardeen black holes and calculate the Lyapunov exponents of the equatorial null circular geodesics. In Section 3, we introduce the two methods used to calculate the massless scalar QNMs. In Section 4, we present the numerical results of the scalar QNMs and discuss the impact of model parameters on them. In Section 5, we verify the connection between the fundamental QNMs and the Lyapunov exponents as well as the shadow radius. The final section is devoted to the conclusion.

	\section{Rotating regular black holes and Lyapunov exponents}\label{section2}
	In this section, we give a brief introduction of the two regular black holes we are interested, i.e., the rotating Hayward black hole and rotating Bardeen black hole proposed in \cite{Bambi:2013ufa}. The metrics of the two black holes can be expressed in the following form
	\begin{equation}\label{RRBH}
		ds^2=-(1-\frac{2f(r)}{\Sigma})dt^2+\frac{\Sigma}{\Delta}dr^2+\Sigma d\theta^2-\frac{4a \sin^2\theta f(r)}{\Sigma}dt d\phi+\frac{\sigma \sin^2\theta}{\Sigma}d\phi^2,
	\end{equation}
	where 
	\begin{eqnarray*}
		\Sigma&=&r^2+a^2\cos^2\theta,\\
		\sigma&=&(r^2+a^2)^2-a^2\Delta\sin^2\theta,\\
		\Delta&=&r^2-2f(r)+a^2,
	\end{eqnarray*}
where $a$ is the rotation parameter of the black holes. To obtain the rotating Hayward black hole and rotating Bardeen black hole, we just take the following replacements respectively,
\bea\label{replace1}
f(r)\to f_h(r)&=&\frac{Mr^4}{r^3+g^3},\\\label{replace2}
f(r)\to f_b(r)&=&\frac{Mr^4}{(r^2+g_*^2)^{3/2}},
\eea
where $g$ and $g_*$ correspond to the deviation parameter and the magnetic charge due to the nonlinear electromagnetic field, respectively.
In the following study, we will often do the calculation with the general function $f(r)$, and then take the replacements for discussing specific models.

Now we consider the Lyapunov exponents of null geodesics in the equatorial plane of the two regular black holes. So we take $\theta=\frac{\pi}{2}$ here. The Lagrangian for a point particle moving in the background of a black hole is
	\begin{equation}
		\mathcal{L}=\frac{1}{2} g_{\mu \nu}\dot{x^\mu}\dot{x^\nu},(\mu,\nu=t,r,\phi),
	\end{equation}
where a dot denotes a derivative with respect to the affine parameter of the null geodesics. The canonical momenta of the particle are defined as
\bea
p_t&=&\frac{\partial \mathcal{L}}{\partial \dot{t}}=g_{tt}\dot{t}+g_{\phi t}\dot{\phi},\label{E}\\
p_\phi&=&\frac{\partial \mathcal{L}}{\partial \dot{\phi}}=g_{\phi \phi}\dot{\phi}+g_{\phi t}\dot{t},\label{L}\\
p_r&=&\frac{\partial \mathcal{L}}{\partial \dot{r}}=g_{rr}\dot{r}.
\eea
Then, the Hamiltonian of the particle in the equatorial plane is
	\begin{equation}\label{eq:H}
		\mathcal{H}=p_t \dot{t}+p_r\dot{r}+p_\phi\dot{\phi}-\mathcal{L}.
	\end{equation}
Since the two regular black holes discussed here have two Killing vectors, $\partial_t$ and $\partial_\phi$, the particle has two corresponding conserved quantities, i.e. the energy $E=-p_t$ and the angular momentum along the rotating axis $L=p_\phi$. According to Eqs.\eqref{E}\eqref{L}, we can obtain
\bea\label{t}
\dot{t}=\frac{E g_{\phi\phi}+g_{\phi t}L}{g_{\phi t}^2-g_{tt}g_{\phi\phi}}=\frac{E g_{\phi\phi}+g_{\phi t}L}{\Delta},\\
\dot{\phi}=\frac{E g_{\phi t}+g_{tt}L}{-g_{\phi t}^2+g_{tt}g_{\phi\phi}}=\frac{E g_{\phi t}+g_{tt}L}{-\Delta}.\label{phi}
\eea
For null geodesics, the Hamiltonian vanishes, i.e.
\bea
2\mathcal{H}=-\frac{L[L(-1+\frac{2f(r)}{\Sigma})-\frac{2a Ef(r)}{\Sigma}]}{\Delta}-\frac{E[-2aLf(r)+E((a^2+r^2)^2-a^2\Delta)]}{\Delta\Sigma}+\frac{\dot{r}^2\Sigma}{\Delta}=0.
\eea
Then we can obtain the radial equation of motion,
\begin{equation}
		r^2\dot{r}^2=a^2E^2-L^2+E^2r^2+\frac{2f(r)(L-aE)^2}{r^2}.
	\end{equation}
Defining $\dot{r}^2=V_r$, the circular null geodesics conditions are $V_r=V_r^\prime=0$ \cite{Bardeen:1972fi}. Suppose the energy and angular momentum of the particle on the circular null geodesics are $ E_c$ and $ L_c$, then the radius $r_c$ of the circular null geodesics satisfies the following two equations,
\bea\label{V}
		a^2E_c^2-L_c^2+E_c^2r_c^2+\frac{2f(r_c)(L_c-aE_c)^2}{r_c^2}=0,\\
\label{Vd}
		\frac{2[E_c^2r_c^4-2(L_c-aE_c)^2f(r_c)+(L_c-aE_c)^2r_cf^\prime(r_c)]}{r_c^3}=0.
	\eea
	Defining the critical impact parameter $d=\frac{L_c}{E_c}$, then the above two equations become
\bea\label{hb0d}
		r_c^2(a^2-d^2+r_c^2)+2(d-a)^2f(r_c)=0,\\
\label{hb1d}
		r_c^4-2(d-a)^2f(r_c)+(d-a)^2r_cf^\prime(r_c)=0.
\eea
For fixed parameters of the black holes, we can obtain $d $ and $r_c$ from the above two equations.

A simple formula for the principal Lyapunov exponent $\lambda$ of the equatorial circular geodesics in terms of the second derivative of the effective potential $V_r$ was derived in \cite{Cardoso:2008bp}, which is 
	\begin{equation}\label{lambda}
		\lambda=\sqrt{\frac{V^{\prime\prime}_r}{2\dot{t}^2}},
	\end{equation}
where a prime stands for a derivative with respect to $r$. Note that the  Lyapunov exponent is calculated with quantities taken at the null circular geodesics. 

	Solving Eq.\eqref{hb1d}, we obtain
	\begin{equation}\label{d}
		d=a\pm\sqrt{\frac{r_c^4}{2 f(r_c)-r_cf^\prime(r_c)}}.
	\end{equation}
The above two solutions of $d$ correspond to the critical impact parameters of the corotating (upper sign) and counterrotating (lower sign) null geodesics. So generally there are also corotating and counterrotating  Lyapunov exponents \cite{Cardoso:2008bp}.
	According to Eq.\eqref{lambda} and the definition of $V_r$, the square of the Lyapunov
	exponent is
	\begin{equation}\label{lambda2}
		\lambda^2=\frac{(aE_c-L_c)[3(aE_c+L_c)r_c^2+(aE_c-L_c)(20f(r_c)+r_c(-8f^\prime(r_c)+r_cf^{\prime\prime}(r_c)))]}{r_c^6\dot{t}^2},
	\end{equation}
	where $\dot{t}=\frac{E_c[(a^2+r_c^2)^2-a^2\Delta]-2aL_cf(r_c)}{\Delta\Sigma}$ according to Eq.\eqref{t}.
	Together with Eqs.\eqref{hb0d}\eqref{hb1d}\eqref{d}\eqref{lambda2}, we can obtain
	\begin{equation}
		\lambda^2=\frac{(a^2-2f(r_c)+r_c^2)^2[r_c^2(3a^2-3d^2+8r_c^2)+(a-d)^2(4f(r_c)+r_c^2f^{\prime\prime}(r_c))]}{[r_c^3(a^2+r_c^2)+2a(a-d)r_cf(r_c)]^2}.
	\end{equation}

With the replacements in Eqs.\eqref{replace1}\eqref{replace2}, 	
we numerically calculate equatorial Lyapunov exponents of rotating Bardeen and Hayward black holes for different values of the parameters $a$, $g$ and $g^*$. 
The corotating and counterrotating Lyapunov exponents of Hayward black holes are listed in Table \ref{LEH} and Table \ref{LEHco}, respectively.
The corotating and counterrotating Lyapunov exponents of Bardeen black holes are listed in Table \ref{LEB} and Table \ref{LEBco}, respectively.
Since we focus on rotating black hole models, so there are constraints on the ranges of the parameters $a$ and $g(g_*)$\cite{Pedrotti:2024znu}. When the values of the parameters don't correspond to black holes, the Lyapunov exponents are omitted in the Tables.

	\begin{table}[h]
		\centering
		\caption{Corotating Lyapunov exponents of the rotating Hayward black holes with different parameters $a$ and $g$.}
		\label{LEH}
		\begin{tabular}{|c|c|c|c|c|c|c|c|c|c|}
			\hline
			\backslashbox{$g$}{$a$} & 0.1 & 0.2 & 0.3 & 0.4 & 0.5 & 0.6 & 0.7 & 0.8 & 0.9 \\
			\hline
			0.1 & 0.192275 & 0.191727 & 0.190645 & 0.18881 & 0.185874 & 0.181242 & 0.173796 & 0.161073 & 0.135731 \\
			\hline
			0.2 & 0.192155 & 0.191581 & 0.190463 & 0.188576 & 0.185563 & 0.180808 & 0.173143 & 0.159958 & 0.133208 \\
			\hline
			0.3 & 0.191827 & 0.191179 & 0.189961 & 0.187932 & 0.184704 & 0.179602 & 0.171312 & 0.156771 & - \\
			\hline
			0.4 & 0.191172 & 0.190377 & 0.188955 & 0.186633 & 0.182958 & 0.177121 & 0.167466 & - & - \\
			\hline
			0.5 & 0.190051 & 0.188995 & 0.187209 & 0.184353 & 0.179849 & - & - & - & - \\
			\hline
			0.6 & 0.188273 & 0.186781 & 0.184374 & 0.180586 & - & - & - & - & - \\
			\hline
			0.7 & 0.185547 & 0.183333 & 0.179862 & - & - & - & - & - & - \\
			\hline
		\end{tabular}
	\end{table}

\begin{table}[h]
		\centering
		\caption{Counterrotating Lyapunov exponents of the rotating Hayward black holes with different parameters $a$ and $g$.}
		\label{LEHco}
		\begin{tabular}{|c|c|c|c|c|c|c|c|c|c|}
			\hline
			\backslashbox{$g$}{$a$} & 0.1 & 0.2 & 0.3 & 0.4 & 0.5 & 0.6 & 0.7 & 0.8 & 0.9 \\
			\hline
			0.1 & 0.192308 & 0.191963 & 0.191453 & 0.190816 & 0.190082 & 0.189271 & 0.188403 & 0.187491 & 0.186544 \\
			\hline
			0.2 & 0.192223 & 0.191891 & 0.19139 & 0.190761 & 0.190033 & 0.189229 & 0.188365 & 0.187456 & 0.186513 \\
			\hline
			0.3 & 0.191992 & 0.191692 & 0.191219 & 0.190611 & 0.189901 & 0.189111 & 0.18826 & 0.187362 & - \\
			\hline
			0.4 & 0.191533 & 0.1913 & 0.190879 & 0.190315 & 0.189641 & 0.188881 & 0.188055 & - & - \\
			\hline
			0.5 & 0.190753 & 0.190635 & 0.190306 & 0.189817 & 0.189203 & 0.188493 & - & - & - \\
			\hline
			0.6 & 0.189532 & 0.189600 & 0.189418 & 0.189045 & 0.188528 & - & - & - & - \\
			\hline
			0.7 & 0.187701 & 0.188059 & 0.188103 & - & - & - & - & - & - \\
			\hline
		\end{tabular}
	\end{table}

\begin{table}[h]
		\centering
		\caption{Corotating Lyapunov exponents of the rotating Bardeen black holes with different parameters $a$ and $g_*$.}
		\label{LEB}
		\begin{tabular}{|c|c|c|c|c|c|c|c|c|c|}
			\hline
			\backslashbox{$g_*$}{$a$} & 0.1 & 0.2 & 0.3 & 0.4 & 0.5 & 0.6 & 0.7 & 0.8 & 0.9 \\
			\hline
			0.1 & 0.192028 & 0.19142 & 0.190255 & 0.188303 & 0.185191 & 0.180278 & 0.172337 & 0.158579 & 0.130104 \\
			\hline
			0.2 & 0.191194 & 0.19038 & 0.188931 & 0.186569 & 0.182835 & 0.176907 & 0.167105 & 0.149151 & 0.104474 \\
			\hline
			0.3 & 0.189652 & 0.18844 & 0.186432 & 0.183245 & 0.178215 & 0.170051 & 0.155734 & 0.12501 & - \\
			\hline
			0.4 & 0.187108 & 0.185191 & 0.182156 & 0.177385 & 0.169689 & 0.156355 & 0.128658 & - & - \\
			\hline
			0.5 & 0.182959 & 0.179743 & 0.174703 & 0.16655 & 0.152245 & 0.121234 & - & - & - \\
			\hline
			0.6 & 0.175815 & 0.169852 & 0.160008 & 0.141743 & 0.0921364 & - & - & - & - \\
			\hline
			0.7 & 0.161534 & 0.14716 & 0.113966 & - & - & - & - & - & - \\
			\hline
		\end{tabular}
	\end{table}

	\begin{table}[h]
		\centering
		\caption{Counterrotating Lyapunov exponents of the rotating Bardeen black holes with different parameters $a$ and $g_*$.}
		\label{LEBco}
		\begin{tabular}{|c|c|c|c|c|c|c|c|c|c|}
			\hline
			\backslashbox{$g_*$}{$a$} & 0.1 & 0.2 & 0.3 & 0.4 & 0.5 & 0.6 & 0.7 & 0.8 & 0.9 \\
			\hline
			0.1 & 0.192028 & 0.19142 & 0.190255 & 0.188303 & 0.185191 & 0.180278 & 0.172337 & 0.158579 & 0.130104 \\
			\hline
			0.2 & 0.191194 & 0.19038 & 0.188931 & 0.186569 & 0.182835 & 0.176907 & 0.167105 & 0.149151 & 0.104474 \\
			\hline
			0.3 & 0.189652 & 0.18844 & 0.186432 & 0.183245 & 0.178215 & 0.170051 & 0.155734 & 0.12501 & - \\
			\hline
			0.4 & 0.187108 & 0.185191 & 0.182156 & 0.177385 & 0.169689 & 0.156355 & 0.128658 & - & - \\
			\hline
			0.5 & 0.182959 & 0.179743 & 0.174703 & 0.16655 & 0.152245 & 0.121234 & - & - & - \\
			\hline
			0.6 & 0.175815 & 0.169852 & 0.160008 & 0.141743 & 0.0921364 & - & - & - & - \\
			\hline
			0.7 & 0.161534 & 0.14716 & 0.113966 & - & - & - & - & - & - \\
			\hline
		\end{tabular}
	\end{table}

	\section{Equations of motion and methods for calculating the QNMs}
In this section, we first consider the equation of motion of massless scalar in rotating Hayward and Bardeen black holes. Then, we introduce the two methods for the calculation of scalar QNMs, i.e. the third-order WKB method and matrix method. 
	\subsection{Klein-Gordon equation}
The equation of motion of a massless scalar perturbation field $\Phi$ is governed by covariant Klein-Gordon equation, which can be written as \cite{Konoplya:2018arm}
	\begin{equation}\label{Klein}
		\frac{1}{\sqrt{-g}}\partial_{\mu}\left(\sqrt{-g} g^{\mu\nu}\partial_{\nu}\Phi\right)=0,
	\end{equation}
	where $\Phi$ is a function of $(t,r,\theta,\phi)$. The perturbation field can be decomposed with the following ansatz
	\begin{equation}\label{Psi}
		\Phi\left(x^{\mu}\right) = e^{-i\omega t} e^{i m\phi} S(\theta) R(r).
	\end{equation}
	Plugging the above ansatz into Eq.\eqref{Klein}, we obtain
	\begin{equation}\label{kgeq1}
		e^{i(m\phi-t\omega)}\frac{(4a^2f(r)^2A(r,\theta)-2\csc^2\theta f(r)B(r,\theta)+\csc^2\theta\Sigma(r,\theta)C(r,\theta))}{\Sigma(r,\theta)(4a^2f(r)^2-2\csc^2\theta f(r)\sigma+\csc^2\sigma(r,\theta)\Sigma(r,\theta))}=0,
	\end{equation}
	where
	\begin{align*}
		A(r,\theta)&=S(R'\Delta(r)+\Delta(r)R'')+R(\cot\theta S'+S''),\\
		B(r,\theta)&=R \left(\sigma (r,\theta ) \left(S''+\cot \theta S'\right)-m \csc ^2\theta S \Sigma (r,\theta )^2 (a \omega  \cos 2 \theta -a \omega +m)\right)+S \sigma (r,\theta ) \left(\Delta (r) R''+\Delta '(r) R'\right),\\
		C(r,\theta)&=R (S \Sigma (r,\theta ) (\omega ^2 \sigma (r,\theta )-m^2 \csc ^2\theta  \Sigma (r,\theta ))+\sigma (r,\theta ) (S''(\theta )+\cot \theta  S'))+S \sigma (r,\theta ) (\Delta (r) R''+\Delta '(r) R').
	\end{align*}
The equation \eqref{kgeq1} is equivalent to
	\begin{equation}
		\frac{\Delta'R'+\Delta R''}{R}+\frac{\cot\theta S'+S''}{S}-a^2\omega^2\sin^2\theta-m^2\csc^2\theta+\frac{a^2m^2+a^4\omega^2+2a^2r^2\omega^2+r^4\omega^2-4am\omega f(r)}{\Delta}=0.
	\end{equation}
Then, the above equation can be divided into the following radial and angular equations 
\bea\label{angular eqaution}
&&S''+\cot\theta S'+(a^2\omega^2\cos^2\theta-m^2\csc^2\theta+E_{lm})S=0,\\
\label{radial equation}
&&\frac{d}{dr}(\Delta R')+\left(\frac{a^2m^2+a^4\omega^2+2a^2r^2\omega^2+r^4\omega^2-4am\omega f(r)}{\Delta}-a^2\omega^2-E_{lm}\right)R=0,
\eea
where $E_{lm}$ is the separation constant, which can be approximated in terms of $(a\omega)$ when $|a\omega|\ll1$ as following \cite{Press:1973zz},
\begin{equation}
	E_{lm}= l(l + 1) + 2a^2\omega^2 \left[ \frac{m^2 + l(l + 1) - 1}{(2l - 1)(2l + 3)} \right] + O[(a\omega)^4].
\end{equation}

In order to solve the quasinormal modes from the above radial equation, we need to choose appropriate physical boundary conditions, i.e. purely ingoing wave at the horizon and purely outgoing wave at spatial infinity. With these boundary conditions, the ansatz for radial function $R(r)$, which will be used later in the numerical calculation, can be written as
\begin{align}\label{radial solution}
	R(r)&=
	\begin{cases}
		e^{i\omega r} \left( \frac{r-r_0}{r-r_1}\right)^{-i\sigma}\left(\frac{r-r_2}{r}\right)^{i\sigma-1}r^{i2M \omega-1}\Psi_R, &\text{Hayward BH},\\
		e^{i\omega r} \left( \frac{r-r_0}{r-r_1}\right)^{-i\sigma}\left(\frac{r-r_2}{r-r_3}\right)^{i\sigma-1}r^{i2M \omega-1}\Psi_R, &\text{Bardeen BH},
	\end{cases}
\end{align}
where $r_0$ and $r_1$ are the event horizon and inner horizon of the black hole, respectively,  $r_2$ and $r_3$ are the negative roots of equation $\Delta(r)=0$, and $\sigma=\frac{a^2\omega+r_0^2\omega-am}{\Delta^\prime(r_0)}$.
Consider the natural boundary conditions for the angular function $S(\theta)$, and we can express the ansatz for the angular function as    
\begin{align}\label{angular solution}
	S(u)&=e^{a\omega u}(1+u)^{|m|/2}(1-u)^{|m|/2}\Phi_S,
\end{align}
where $u=\cos\theta$. 

Defining the tortoise coordinate $r_*$ by $dr/dr_*=\Delta/(r^2+a^2)$  and a new radial function $\psi=\sqrt{r^2+a^2}R(r)$, 
the radial equation of motion \eqref{radial equation} can be written as a $\text{Schr\"odinger-like}$  equation in tortoise coordinate,
\begin{align}\label{radialequationt}
	\frac{d^2\psi(r_*)}{dr_*^2} &+ \left( \frac{-\Delta (r) \left(a^2 \omega ^2+E\right) + a^2 m^2 + \omega ^2 \left(a^2+r^2\right)^2 - 4 a m \omega f(r)}{\left(a^2+r^2\right)^2} \right. \nonumber \\
	&\left. - \frac{\Delta (r) \left(r \left(a^2+r^2\right) \Delta '(r) + \left(a^2-2 r^2\right) \Delta (r)\right)}{\left(a^2+r^2\right)^4} \right) \psi(r_*)=0,
\end{align}
which is useful in the calculation of QNMs with the WKB method.
 
\subsection{Methods for calculating QNMs}
\subsubsection{The third-order WKB method}
Using the third-order WKB method, Iyer and Will computed the QNMs of Schwarzschild black hole \cite{1987Black}. This method was also applied to calculate the QNMs of Reissner-Nordstrom black hole \cite{Iyer:1986nq} and Kerr black hole \cite{Seidel:1989bp}.

We will use the third-order WKB method to calculate the QNMs of rotating regular black holes with small rotation parameters. Here we give a brief introduction of this method. 
The radial equation of motion \eqref{radialequationt} in the rotating regular black holes can be rewritten as the following form of a second-order ordinary differential equation, 
 \begin{equation}
 	\frac{d^{2}\psi(x)}{d x^{2}} + Q(x)\psi(x) = 0,
 \end{equation}
which is similar to the one-dimensional potential barrier problem in quantum mechanics.The function $ Q(x) $ can be expressed in the form $Q(x) = \omega^2 - V(x) $, then the QNMs can be determined by the following equation \cite{Seidel:1989bp},
 	\begin{equation}\label{omega2}
 	\omega^{2} = V_{0}(\omega) + \left[-2 V_{0}^{\prime\prime}(\omega)\right]^{1/2} \Lambda(\omega) - i\left(n + \frac{1}{2}\right) \left[-2 V_{0}^{\prime\prime}(\omega)\right]^{1/2} [1 + \Omega(\omega)],
 \end{equation}
 where
 \begin{equation}
 	\Lambda(\omega) = \frac{1}{\left[-2 V_{0}^{\prime\prime}(\omega)\right]^{1/2}} \left[ \frac{1}{8} \left( \frac{V_{0}^{(4)}}{V_{0}^{\prime\prime}} \right) \left( \frac{1}{4} + \alpha^{2} \right) - \frac{1}{288} \left( \frac{V_{0}^{\prime\prime\prime}}{V_{0}^{\prime\prime}} \right)^{2} \left( 7 + 60 \alpha^{2} \right) \right],
 \end{equation}
 
 \begin{align}
 	\Omega(\omega)=&\frac{1}{-2V_0^{\prime\prime}}[\frac{5}{6912}\left(\frac{V_0^{\prime\prime\prime}}{V_0^{\prime\prime}}\right)^4\left( 77 + 188 \alpha^{2} \right) - \frac{1}{384} \left( \frac{V_{0}^{\prime\prime\prime 2} V_{0}^{(4)}}{V_{0}^{\prime\prime 3}} \right) \left( 51 + 100 \alpha^{2} \right) + \frac{1}{2304} \left( \frac{V_{0}^{(4)}}{V_{0}^{\prime\prime}} \right)^{2} \left( 67 + 68 \alpha^{2} \right)\nonumber\\
 	&+ \frac{1}{288} \left( \frac{V_{0}^{\prime\prime\prime} V_{0}^{(5)}}{V_{0}^{\prime\prime 2}} \right) \left( 19 + 28 \alpha^{2} \right) - \frac{1}{288} \left( \frac{V_{0}^{(6)}}{V_{0}^{\prime\prime}} \right) \left( 5 + 4 \alpha^{2} \right)],
 \end{align}
  
 \begin{equation}
 	\alpha=n+\frac{1}{2},
 \end{equation}
 
 \begin{equation}
 	n = \left\{
 	\begin{array}{l}
 		0, 1, 2, \ldots, \text{Re}(\omega) > 0 \\
 		-1, -2, -3, \ldots, \text{Re}(\omega) < 0
 	\end{array}
 	\right.
 \end{equation}
 The primes represent derivatives with respect to $r_*$. The subscript 0 indicates that the derivatives are to be evaluated at the value of $r_*$, for which the potential $V$ obtains its maximum. We use $r_e$ to indicate the value of $r$ when $V$ obtains its maximum.  
  When $|a\omega|\ll1$, we can expand the potential $V(r)$ and $r_e$ as
 \bea\label{V6}
 	V = V_{s} +(a\omega)V_{1} +(a\omega)^{2} V_{2} +(a\omega)^{3}V_{3} +(a\omega)^{4}V_{4}  +(a\omega)^{5}V_{5} +(a\omega)^{6}V_{6} + \cdots,\\
\label{r0}
 	r_e =  r_s + r_1(a\omega) + r_2(a\omega)^2 + r_3(a\omega)^3 + r_4(a\omega)^4 
 	+ r_5(a\omega)^5 + r_6(a\omega)^6 + \cdots = r_s + \Sigma,
 \eea
 where $V_s$ is the potential governing the perturbation of the corresponding nonrotating regular black hole, $\{V_1,V_2,..\}$ are known expansion coefficients as functions of $r$, 
$r_s$ is the value of $r$ such that $V_s(r )$ reaches its maximum, and $\{r_1,r_2,..\}$ are expansion coefficients to be determined.   
Since $r_e$ is the value where $V(r)$ reaches its maximum, we obtain
 	\begin{align}
 	0 = & V^{\prime}\left(r_{e}\right) 
 	=  V^{\prime}\left(r_{s}+\Sigma\right) \nonumber\\
 	= & V^{\prime}\left(r_{s}\right) + \Sigma V^{\prime\prime}\left(r_{s}\right) + \frac{1}{2}\Sigma^{2} V^{\prime\prime\prime}\left(r_{s}\right) 
 	 + \frac{1}{6}\Sigma^3 V^{(4)}\left(r_s\right) + \frac{1}{24}\Sigma^4 V^{(5)}\left(r_s\right)\nonumber \\
 	& + \frac{1}{120}\Sigma^{5} V^{(6)}\left(r_{s}\right) + \frac{1}{720}\Sigma^{6} V^{(7)}\left(r_{s}\right)+\cdots.
 \end{align}
Plugging Eqs.\eqref{V6}\eqref{r0} into the above equation, and expand all quantities through order $(a\omega)^6$,  we can determine $\{r_1,r_2,..\}$ perturbatively. 
Then, substituting them into Eq.\eqref{omega2}, we can numerically compute the QNMs frequency $\omega$ with fixed other model parameters, such as $M,a,g(g_*), n, l$ and $m$. 

\subsubsection{The Matrix Method}
The matirx method for calculating the QNMs has been proposed for Schwarzschild BH \cite{Lin:2016sch} and Kerr BH \cite{Lin:2017oag}. One advantage of this method is that we need not assume $|a\omega|\ll1$. Here we give a brief introduction of this method for the rotating regular black holes. 
Begin with the ansatz solutions of the radial and angular equations, Eqs.\eqref{radial solution}\eqref{angular solution}, 
we introduce the coordinate transformations from $r$ and $u$ to $x$ and $z$, and transform the wave
functions from$\Psi_R$ and $\Phi_S$ to $\chi_R$ and $\chi_S$ as follows:
\begin{equation}\label{cordinate transformation}
	\begin{aligned}
   x &= 1 - \frac{r_0}{r},~ z = \frac{1+u}{2},\\
		\chi_R &= x(x-1)\Psi_R,~ \chi_S = z(z-1)\Phi_S. 
	\end{aligned}
\end{equation}
Then the ranges of the new coordinates are $0\leq x\leq1,0\leq z\leq1$, and the new wave functions satisfy the following boundary conditions
\begin{equation}
	\chi_R(0) = \chi_R(1) = \chi_S(0) = \chi_S(1) = 0.
\end{equation}
Substituting Eqs.\eqref{radial solution}\eqref{angular solution}\eqref{cordinate transformation} into Eqs.\eqref{angular eqaution}\eqref{radial equation}, the equations satisfied by the new wave functions are
\begin{equation}
	\begin{aligned}\label{juzheneq}
		\tau_R(x) \chi_R''(x) + \lambda_R(x) \chi_R'(x) + s_R(x) \chi_R(x) &= 0, \\
		\tau_S(z) \chi_S''(z) + \lambda_S(z) \chi_S'(z) + s_S(z) \chi_S(z) &= 0,
	\end{aligned}
\end{equation}
where \begin{equation}
	\begin{aligned}
	\tau_R(x)&=(x-1)^8 x^2 \Delta^2,\\
	\lambda_R(x)&=-i(x-1)^6 x\Delta
	(2(-i + \sigma + x^2(-i + \sigma + \omega) - x(-2i + 2\sigma + \omega + r_0\omega)) \Delta+ i r_0 x \Delta'),\\
	s_R(x)&= {r_0}^2 x^2 (r_0^4 \omega^2 + a^4 (x-1)^4 \omega^2 + a^2 (x-1)^2 (m^2 (x-1)^2 + 2 r_0^2 \omega^2)) - 4 a m r_0^2 (x-1)^4 x^2 \omega f \\
	&-(x-1)^4 (-2 - 3 i\sigma + \sigma^2 - 2 x^3 (-2 i + \sigma + \omega) (-2 i + 2 \sigma + \omega + r_0 \omega) - 2 x (-i + \sigma) (-4 i + 2 \sigma + \omega + r_0 \omega) +\\
	& x^4 (-2 + \sigma^2 - 3 i \omega + \omega^2 + \sigma (-3 i + 2 \omega)) + x^2 (-12 + 6 \sigma^2 - i (7 + 6 r_0) \omega + (1 + r_0)^2 \omega^2 + 2 \sigma (-9 i + (3 + 2 r_0) \omega))) \Delta^2\\
	&-r_0 (x-1)^4 x \Delta (r_0 x (E_{lm} + a^2 \omega^2) + i (-i + \sigma + x^2 (-2 i + \sigma + \omega) - x (-3 i + 2 \sigma + \omega + r_0 \omega)) \Delta'),\\
	\tau_S(z)&=(z-1)^2z^2,\\
	 \lambda_S(z)&=z (z-1) (1+m(2z-1)+4az^2\omega-2z(1+2a\omega)),\\
	 s_S(z)&=1+m^2(z-1)z-4a z^3\omega-z^2(-2+E_{lm}-6a\omega+a^2\omega^2)\\
	 &+z(-2+E_{lm}-2a \omega+a^2\omega^2)+m(-1+4az^3\omega-3z^2(1+2a\omega)+z(3+2a\omega)).
	\end{aligned}
\end{equation}
The next step in the matrix method is to transform the new radial and angular differential equations \eqref{juzheneq} to matrix equations \cite{Lin:2016sch,Lin:2017oag,Huang:2025wom}.
We sketch this procedure with the radial equation. Choose $N$ different points $\{x_1,x_2,...,x_N\}$ in the range $[0,1]$ and take $x_1=0,x_N=1$, then we expand the function $\chi_R$
around $x_j$ to express the values of  $\chi_R$ at other points with Taylor formula up to $(N-1)-$th order. Thus, we can express $\chi'_R(x_j),\chi''_R(x_j)$ as linear combination of values of $\chi_R$ at the $N$ points. Then, the differential radial equation can be rewritten as a matrix equation about the values of $\chi_R$ at the $N$ points\cite{Huang:2025wom}, which takes the following form 
\bea
\bar{\mathcal{M}}_R \mathcal{F}_R = 0,
\eea      
where the elements of $\bar{\mathcal{M}}_R$ depend on the values of $\tau_R,\lambda_R,s_R$ at the $N$ points, and $\mathcal{F}_R=\{\chi_R(x_1),\chi_R(x_2),\cdots,\chi_R(x_N)\}^T$.
Considering the boundary conditions $\chi_R(0)=\chi_R(1)=0$, we can finally obtain the radial matrix equation 
\begin{equation}\label{radialMteq}
	\mathcal{M}_R \mathcal{F}_R = 0,
\end{equation}
where
\begin{equation}
	\left(\mathcal{M}_R\right)_{k, i} = 
	\begin{cases} 
		\delta_{k, i}, & k=1 \text{ or } N, \\
		\left(\bar{\mathcal{M}}_R\right)_{k, i}, & k=2,3,\ldots, N-1.
	\end{cases}
\end{equation}
Similarly, for the new angular equation, we also have the corresponding matrix equation  
\bea\label{angularMteq}
 \mathcal{M}_S \mathcal{F}_S = 0,
\eea
where
\begin{equation}
	\left(\mathcal{M}_S\right)_{k, i} = \begin{cases} 
		\delta_{k, i}, & k=1 \text{ or } N, \\
		\left(\bar{\mathcal{M}}_S\right)_{k, i}, & k=2,3,\ldots, N-1.
	\end{cases}
\end{equation}
The separation constant $E_{lm}$ and QNM frequency $\omega$ are determined by the requirement that the homogeneous matrix equations \eqref{radialMteq}\eqref{angularMteq} have nonzero solutions, i.e.
\begin{equation}
	\det(\mathcal{M}_R)=0,~ \det(\mathcal{M}_S)=0.
\end{equation}
In our numerical calculation, we take $N=20$ and the mass parameter of the black holes $M=1$.

\section{Numerical results of the QNMs} 

\begin{table}
\caption{Comparison of fundamental QNMs of massless scalar field in Kerr black hole calculated with different methods. $\omega_{\text{MM}}$ and $\omega_{\text{WKB}}$ are calculated with our matrix method and WKB method, respectively. $\omega_{\text{CFM}}$ is adopted from Ref.\cite{Konoplya:2006br}. Other parameters are chosen as $M=1$, $l= 1$, $m = 1$.}
\label{comparison}
\begin{tabular}{ccccccc}
	\toprule
	\multicolumn{1}{c}{} & \multicolumn{2}{c}{$\omega_{\text{CFM}}$}& \multicolumn{2}{c}{$\omega_{\text{MM}}$}& \multicolumn{2}{c}{$\omega_{\text{WKB}}$}\\
	$a$ & $\text{Re}(\omega)$ & $-\text{Im}(\omega)$
	&  $\text{Re}(\omega)$ & $-\text{Im}(\omega)$& $\text{Re}(\omega)$ & $-\text{Im}(\omega)$\\
	\midrule
	0 & 0.292936 & 0.097660 \;\;&0.292937&0.097659\;\;&0.291114& 0.098001\\
	0.1 & 0.301045 & 0.097547\;\;&0.301046&0.097546\;\;& 0.299289 & 0.097950 \\
	0.2 & 0.310043 & 0.097245\;\;&0.310044&0.097244\;\;&0.308474&0.097780 \\
	0.3 & 0.320126 & 0.096691\;\; &0.320128&0.096691\;\;& 0.318891& 0.097435  \\
	0.4 & 0.331567 & 0.095792 \;\;&0.331568&0.095791\;\; & 0.330834 &0.096808 \\
	0.5 & 0.344753 & 0.094395\;\; &0.344755&0.094394\;\;& 0.344682&  0.095658\\
	0.6 & 0.360285 & 0.092243\;\; &0.360286&0.092243\;\;&  0.360789& 0.093195\\
	0.7 & 0.379159 & 0.088848\;\;&0.379160 &0.088849\;\;& 0.377703 & 0.086979\\
	0.8 & 0.403273 & 0.083132\;\;&0.403274 &0.083133 \;\;&- & - \\
	\bottomrule
\end{tabular}
\end{table}

In order to validate the two methods used in this work, we compare the fundamental ($n=0$) massless scalar QNMs in Kerr black hole calculated with our methods and that calculated with Leaver's continued fraction method \cite{Konoplya:2006br} in Table \ref{comparison}. We can see that the results calculated with matrix method have enough accuracy, while the results from WKB method are less accurate. When $a=0.8$, there is a significant deviation in the result from WKB method and we don't show it. Due to the higher accuracy of the matrix method, the results presented in the following are calculated with this method.  

\subsection{Rotating Hayward black holes}

In this subsection we present the numerical results of the scalar QNMs in the rotating Hayward black holes. We calculate the fundamental ($n=0$) QNMs and higher overtone QNMs with $n=1$, and show the impacts of the model parameters on the values of the QNMs.  

\begin{figure}
	\centering
	\begin{minipage}{0.45\textwidth}
		\includegraphics[width=\linewidth]{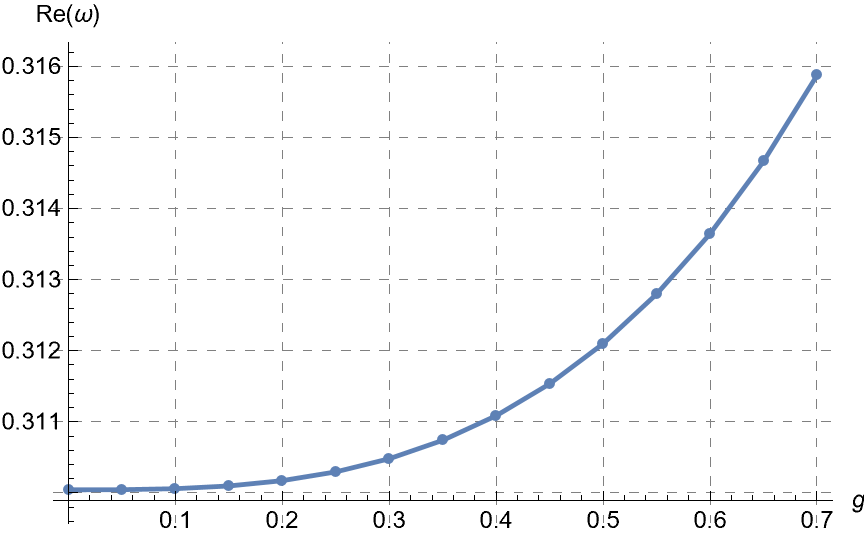}
		
	\end{minipage}\hfill
	\begin{minipage}{0.45\textwidth}
		\includegraphics[width=\linewidth]{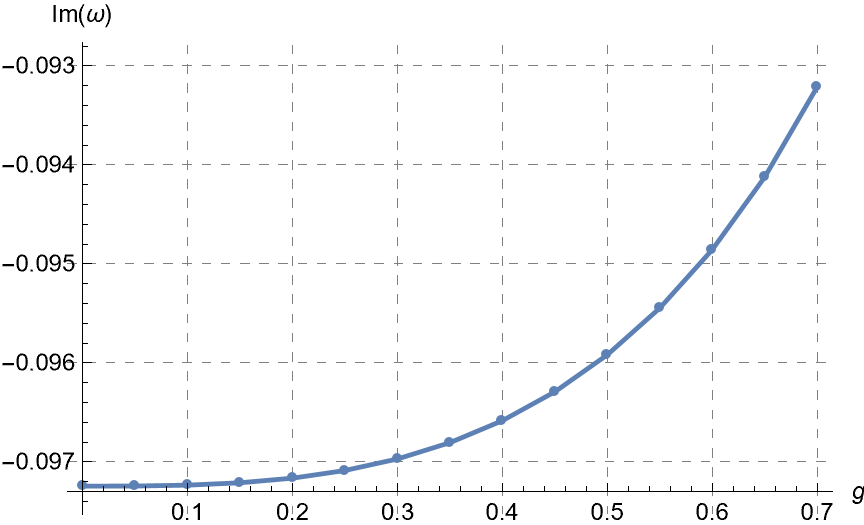}
		
	\end{minipage}
	\caption{The fundamental QNMs of massless scalar field in rotating Hayward black holes with different values of $g$. Left: the real parts of the QNMs. Right: the imaginary parts of the QNMs. Other parameters are chosen as $M=1,a=0.2,l=1,m=1$.}\label{n0HayVg}
\end{figure}

In Fig.\ref{n0HayVg}, we plot the fundamental QNMs of the massless scalar field in the rotating Hayward black holes. Impact of the parameter $g$ on the real and imaginary parts of the QNMs are shown. 
Although the real and imaginary parts of the fundamental QNMs monotonically increase as $g$ increases, the parameter $g$ has minor impacts on the values of the real and imaginary parts of the fundamental QNMs. Compared with the Kerr black hole ($g=0$) case, the real part for the case with $g=0.7$ has about $1.9\%$ increase, and the imaginary part has about $4.1\%$ increase. 

\begin{figure}
	\centering
	\begin{minipage}{0.45\textwidth}
		\includegraphics[width=\linewidth]{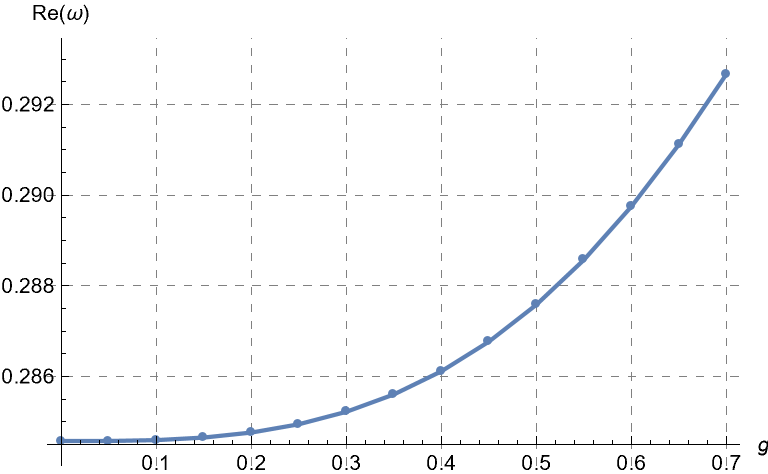}
		
	\end{minipage}\hfill
	\begin{minipage}{0.45\textwidth}
		\includegraphics[width=\linewidth]{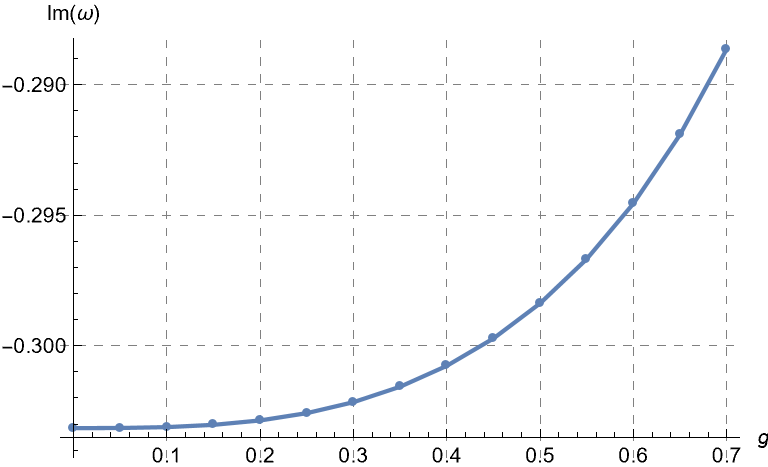}
		
	\end{minipage}
	\caption{The $n=1$ QNMs of massless scalar field in rotating Hayward black holes with different values of $g$. Left: the real parts of the QNMs. Right: the imaginary parts of the QNMs. Other parameters are chosen as $M=1,a=0.2,l=1,m=1$.}
	\label{hayward_g}
\end{figure}

The impacts of the parameter $g$ on the real and imaginary parts of the QNMs with overtone number $n=1$ are shown in Fig.\ref{hayward_g}. Similar to the fundamental QNMs, 
the parameter $g$ has minor impacts on the values of the real and imaginary parts of the $n=1$ QNMs. Compared with the Kerr black hole ($g=0$) case, the real part for the case with $g=0.7$ has about $2.8\%$ increase, and the imaginary part has about $3.7\%$ increase.

\begin{figure}
	\centering
	\begin{minipage}{0.45\textwidth}
		\includegraphics[width=\linewidth]{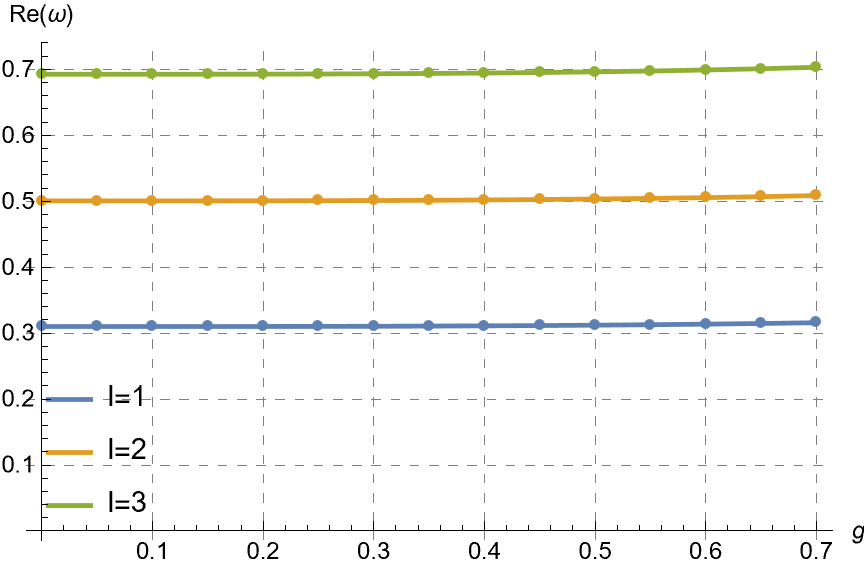}
	\end{minipage}\hfill
	\begin{minipage}{0.45\textwidth}
		\includegraphics[width=\linewidth]{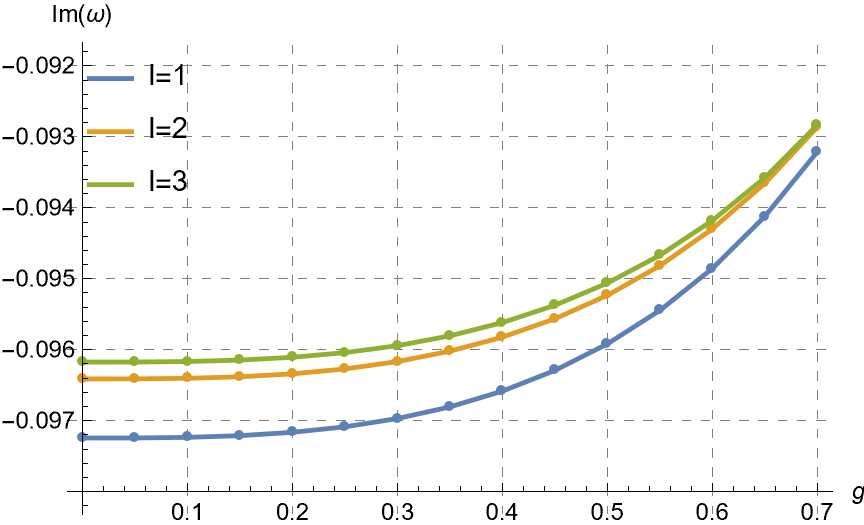}
	\end{minipage}
	\caption{Fundamental QNMs in rotating Hayward black holes, with $M=1, a=0.2, m=1$ for different values of $g$ and $l$.}
	\label{hayward_l123_n0}
\end{figure}

\begin{figure}
	\centering
	\begin{minipage}{0.45\textwidth}
		\includegraphics[width=\linewidth]{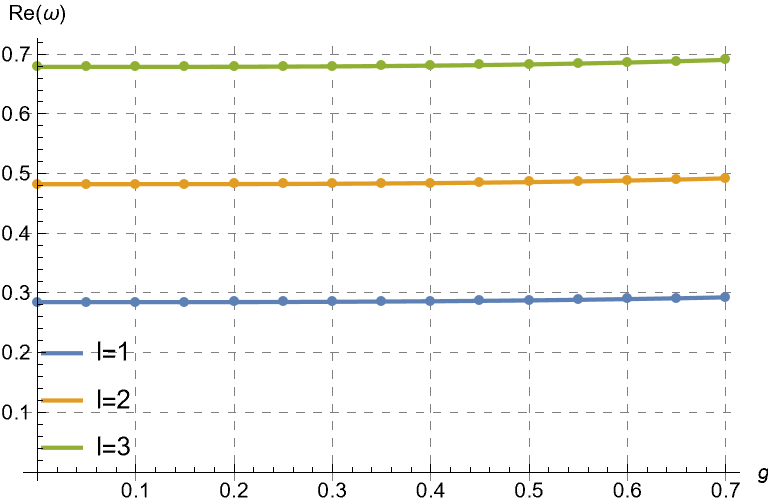}
	\end{minipage}\hfill
	\begin{minipage}{0.45\textwidth}
		\includegraphics[width=\linewidth]{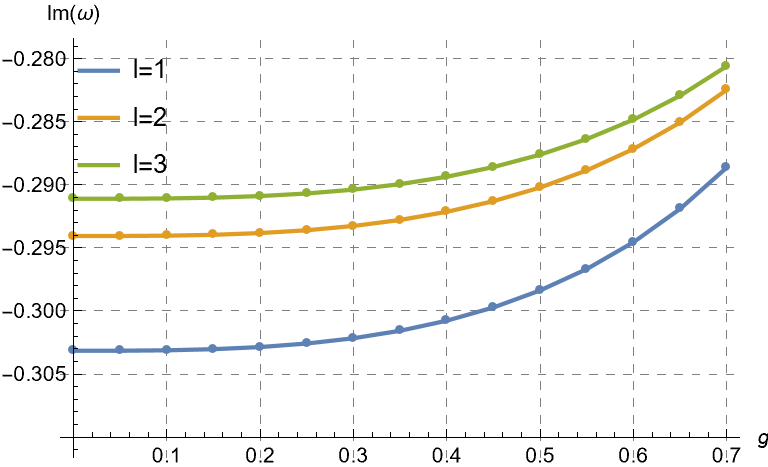}
	\end{minipage}
	\caption{$n=1$ QNMs in rotating Hayward black holes, with $M=1, a=0.2, m=1$ for different values of $g$ and $l$.}
	\label{hayward_l123}
\end{figure}

 In Fig.\ref{hayward_l123_n0}, we show the fundamental QNMs in rotating Hayward black holes for different values of $g$ and $l$. We can see that multipole number $l$ has a significant impact on the real parts of the QNMs while a small impact on the imaginary parts. In Fig.\ref{hayward_l123},  we show the $n=1$ QNMs in rotating Hayward black holes for different values of $g$ and $l$. From Fig.\ref{hayward_l123_n0} and Fig.\ref{hayward_l123}, we notice that, contrary to the multipole number $l$, the overtone number $n$ has a significant impact on the imaginary parts of the QNMs while small impact on the real parts. Furthermore, for all cases ($l=1,2,3$) shown here, the real parts of the QNMs with overtone $n=1$ are a little smaller than that with $n=0$.

\subsection{Rotating Bardeen black holes}

In this subsection we present the numerical results of the scalar QNMs in the rotating Bardeen black holes. We calculate the fundamental ($n=0$) QNMs and higher overtone QNMs with $n=1$, and show the impacts of the model parameters on the values of the QNMs.  

\begin{figure}
	\centering
	\begin{minipage}{0.45\textwidth}
		\includegraphics[width=\linewidth]{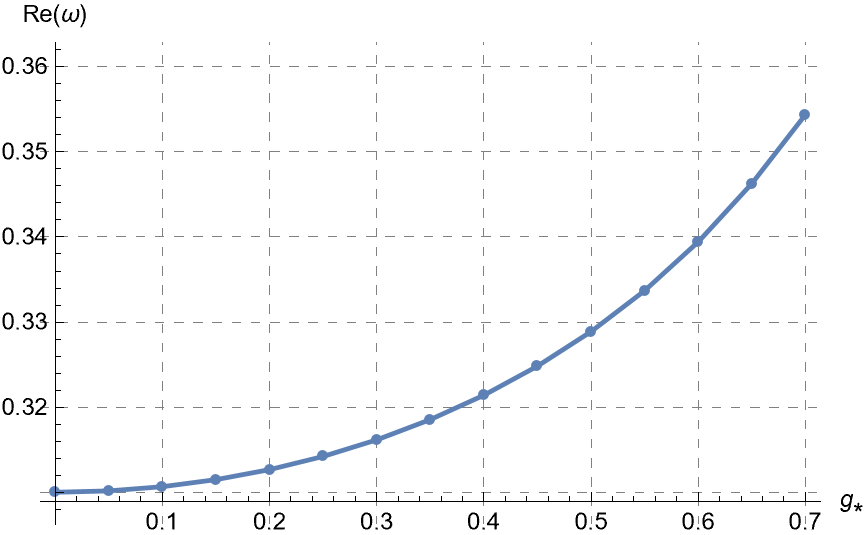}
	\end{minipage}\hfill
	\begin{minipage}{0.45\textwidth}
		\includegraphics[width=\linewidth]{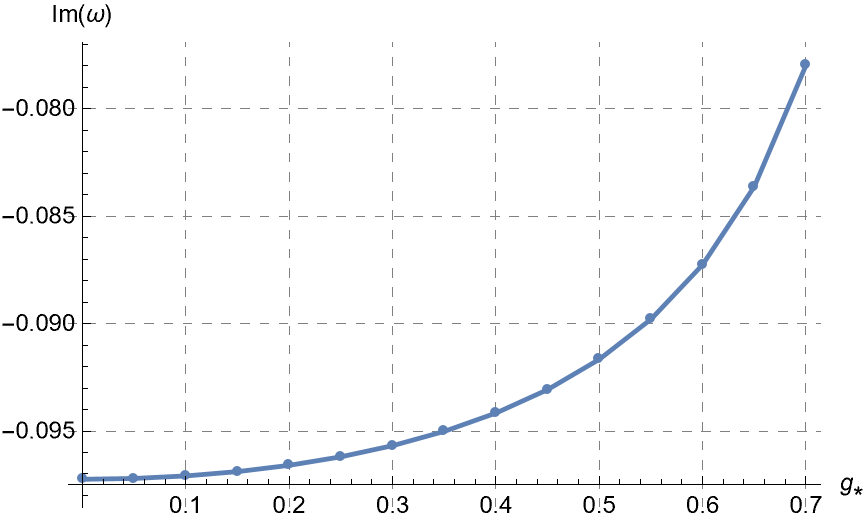}
	\end{minipage}
	\caption{The fundamental QNMs of massless scalar field in rotating Bardeen black holes with different values of $g_*$. Left: the real parts of the QNMs. Right: the imaginary parts of the QNMs. Other parameters are chosen as $M=1,a=0.2,l=1,m=1$.}\label{bardeen_gn0}
\end{figure}

In Fig.\ref{bardeen_gn0}, we plot the fundamental QNMs of the massless scalar field in the rotating Bardeen black holes. Impact of the parameter $g_*$ on the real and imaginary parts of the QNMs are shown. The real and imaginary parts of the fundamental QNMs monotonically increase as $g_*$ increases. Compared with the Hayward black hole case, we notice that the parameter $g_*$ has relatively much larger impacts on the values of the real and imaginary parts of the fundamental QNMs. 
The real part of the QNM with $g_*=0.7$ increases about $13\%$ over that of the Kerr black hole ($g_*=0$), and the imaginary part increases about $19\%$.

\begin{figure}
	\centering
	\begin{minipage}{0.45\textwidth}
		\includegraphics[width=\linewidth]{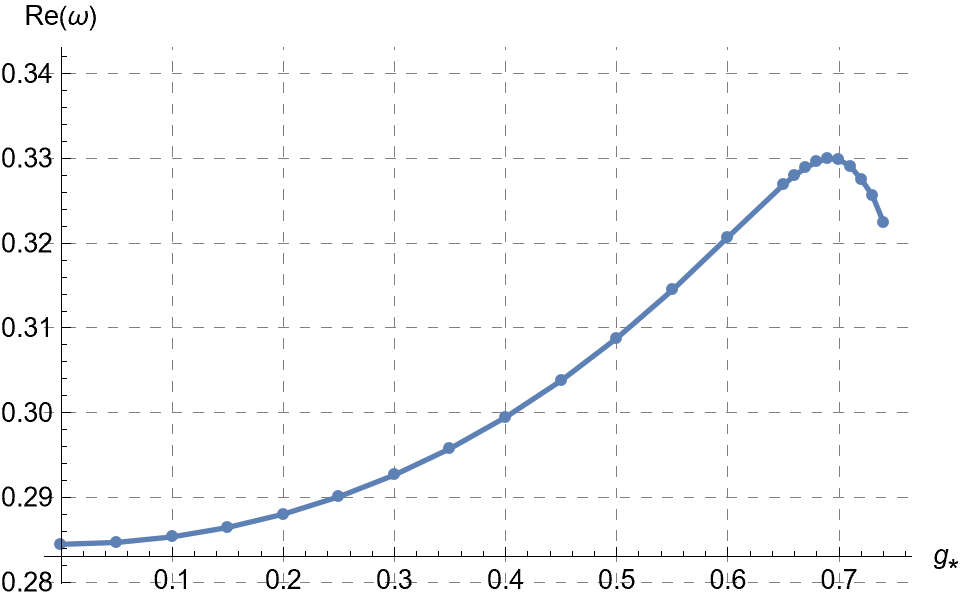}
	\end{minipage}\hfill
	\begin{minipage}{0.45\textwidth}
		\includegraphics[width=\linewidth]{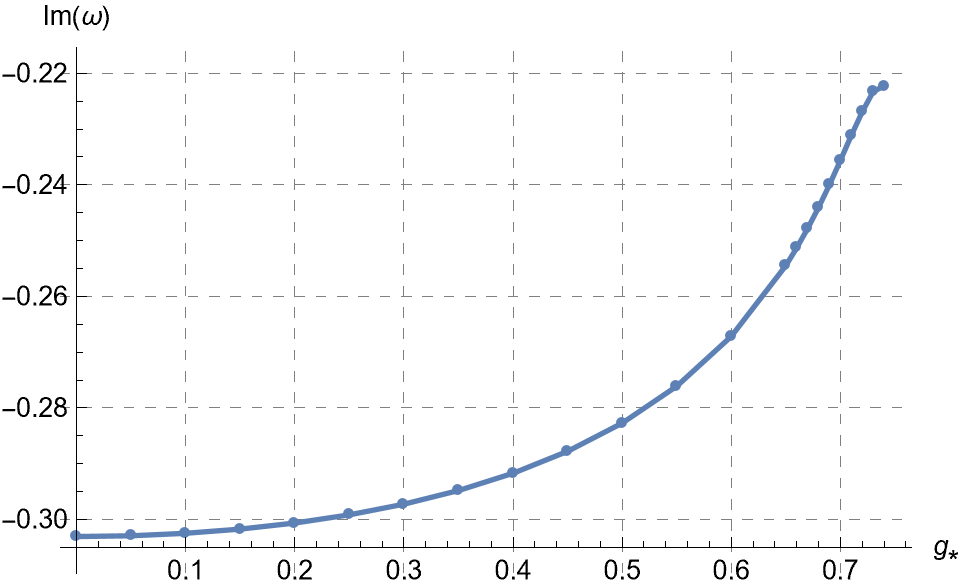}
	\end{minipage}
	\caption{The $n=1$ QNMs of massless scalar field in rotating Bardeen black holes with different values of $g_*$. Left: the real parts of the QNMs. Right: the imaginary parts of the QNMs. Other parameters are chosen as $M=1,a=0.2,l=m=1$.}
	\label{bardeen_g}
\end{figure}

The impacts of the parameter $g_*$ on the real and imaginary parts of the $n=1$ QNMs are shown in Fig.\ref{bardeen_g}. Similar to the fundamental QNMs, the imaginary part of the QNM still monotonically increases as $g_*$ increases. However, 
the real part of the QNM first increases as $g_*$ increases from zero, and when $g_*$ is close to its maximum value 0.744 the real part reaches a maximum and then has a small decrease.    
The parameter $g_*$ also has significant impacts on the values of the real and imaginary parts of the $n=1$ QNMs. Compared with the Kerr black hole ($g_*=0$) case, the real part of the QNMs with $g_*=0.7$ increases about $16\%$ , and the imaginary part increases about $19\%$. 

\begin{figure}
	\centering
	\begin{minipage}{0.45\textwidth}
		\includegraphics[width=\linewidth]{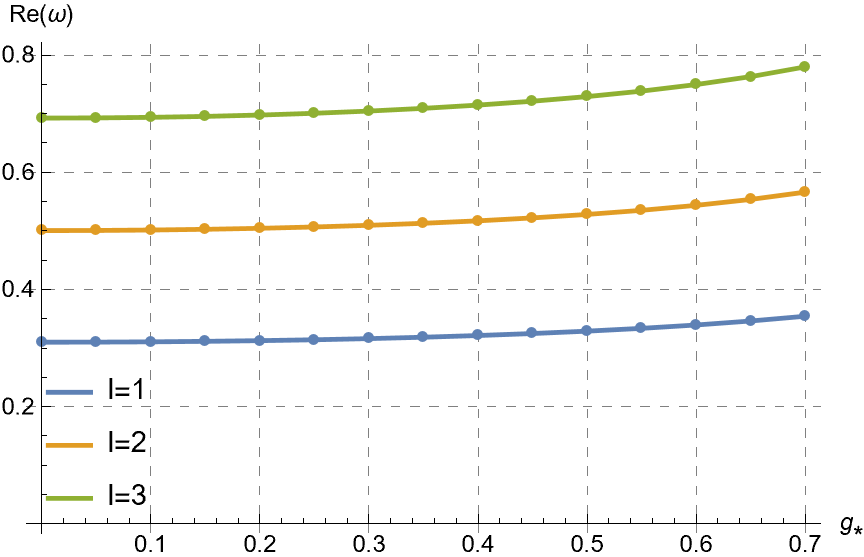}
	\end{minipage}\hfill
	\begin{minipage}{0.45\textwidth}
		\includegraphics[width=\linewidth]{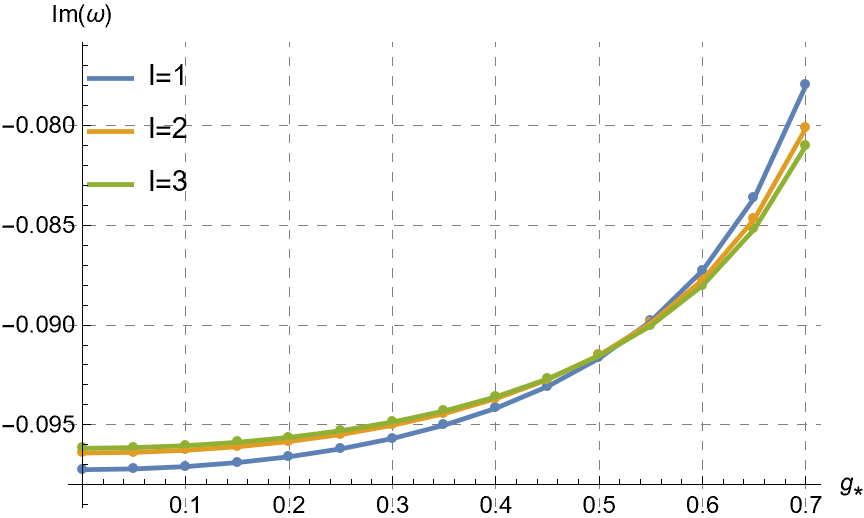}
	\end{minipage}
	\caption{Fundamental scalar QNMs in rotating Bardeen black holes, with $M=1, a=0.2, m=1$ for different values of $g_*$ and $l$}
	\label{bardeen_l123_n0}
\end{figure}

\begin{figure}
	\centering
	\begin{minipage}{0.45\textwidth}
		\includegraphics[width=\linewidth]{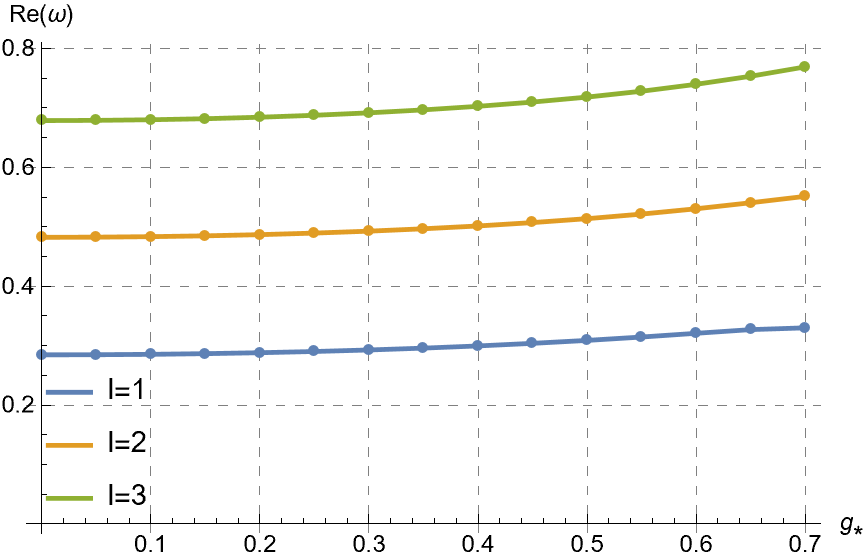}
	\end{minipage}\hfill
	\begin{minipage}{0.45\textwidth}
		\includegraphics[width=\linewidth]{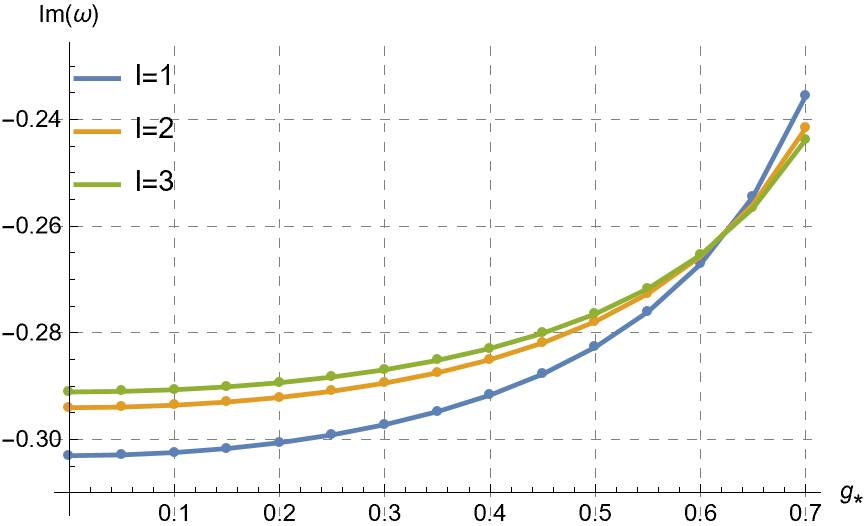}
	\end{minipage}
	\caption{$n=1$ scalar QNMs in rotating Bardeen black holes, with $M=1, a=0.2, m=1$ for different values of $g_*$ and $l$}
	\label{bardeen_l123}
\end{figure}

 In Fig.\ref{bardeen_l123_n0}, we show the fundamental QNMs in rotating Bardeen black holes for different values of $g_*$ and $l$. We can see that multipole number $l$ has a significant impact on the real parts of the QNMs while a relatively small impact on the imaginary parts. In Fig.\ref{bardeen_l123},  we show the $n=1$ QNMs in rotating Bardeen black holes for different values of $g_*$ and $l$. The multipole number $l$ has similar impacts on the real and imaginary parts of the QNMs as the $n=0$ case. Particularly, 
there is a crossing for the imaginary parts of different multipole number $l$ as the increase of $g_*$. There is no such crossing in the Hayward black hole case.  
 From Fig.\ref{bardeen_l123_n0} and Fig.\ref{bardeen_l123}, we also notice that the overtone number $n$ has a significant impact on the imaginary parts of the QNMs while small impact on the real parts. Furthermore, for all cases ($l=1,2,3$), the real parts of the QNMs with overtone $n=1$ are a little smaller than that with $n=0$, which is similar to the Hayward black hole case.


\begin{figure}
	\centering
	\begin{minipage}{0.45\textwidth}
		\includegraphics[width=\linewidth]{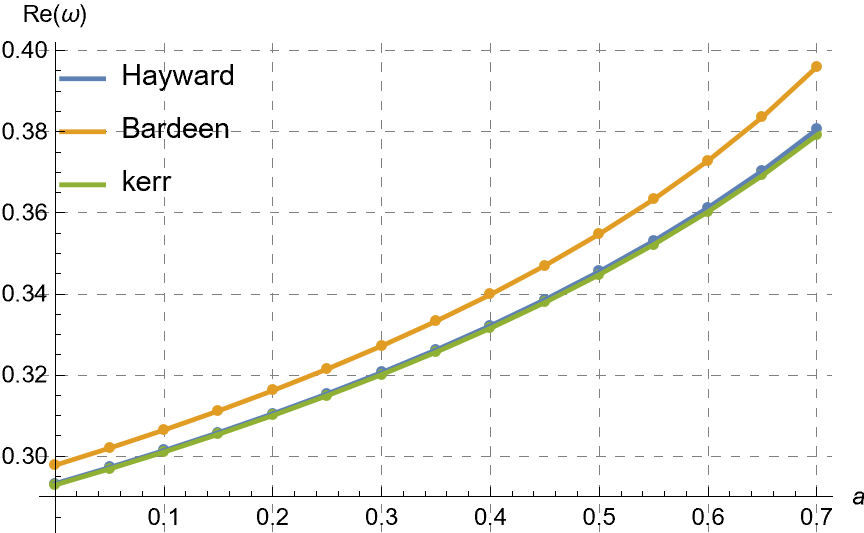}
	\end{minipage}\hfill
	\begin{minipage}{0.45\textwidth}
		\includegraphics[width=\linewidth]{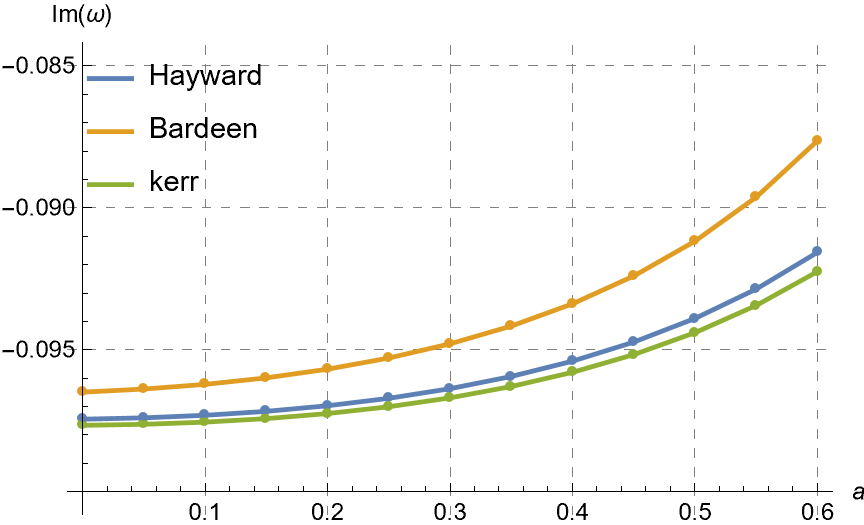}
	\end{minipage}
	\caption{Impact of $a$ on the \( n = 0 \) scalar QNMs in rotating Bardeen and Hayward black holes with \(g=g_* = 0.3 \), \(l= m = 1 \). The Kerr black hole case is also shown.}
	\label{kerr_n0}
\end{figure}

\begin{figure}
	\centering
	\begin{minipage}{0.45\textwidth}
		\includegraphics[width=\linewidth]{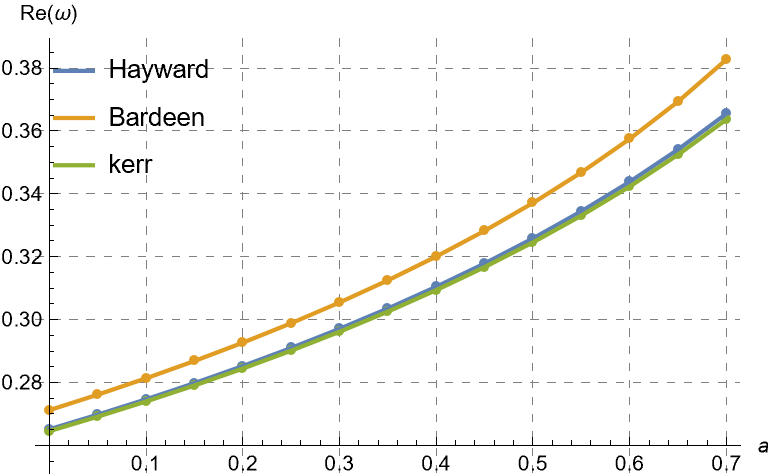}
	\end{minipage}\hfill
	\begin{minipage}{0.45\textwidth}
		\includegraphics[width=\linewidth]{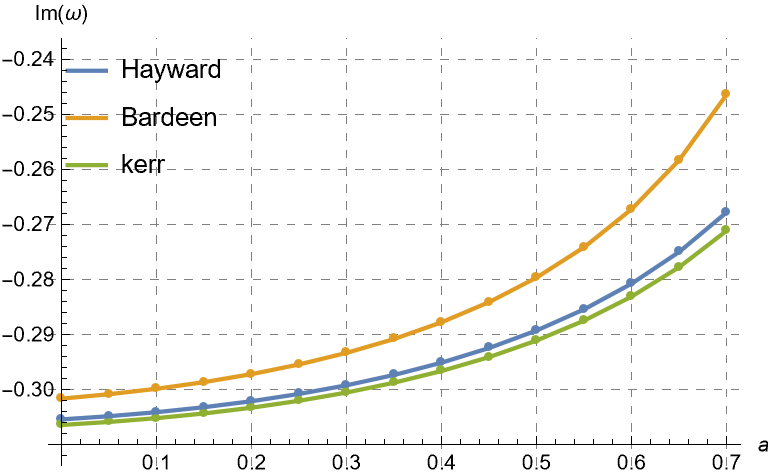}
	\end{minipage}
	\caption{Impact of $a$ on the \( n = 1 \) scalar QNMs in rotating Bardeen and Hayward black holes with \(g=g_* = 0.3 \), \(l= m = 1 \). The Kerr black hole case is also shown.}
	\label{kerr}
\end{figure}

In Fig.\ref{kerr_n0}, the impact of the rotation parameter $a$ on the fundamental scalar QNMs in Kerr, rotating Hayward and Bardeen black holes is shown. In all cases, the real and imaginary parts of the QNMs increase as the increase of rotation parameter $a$. We can notice that the difference of the values of the QNMs between the rotating Hayward black hole case and Kerr case is very small. And the QNMs in rotating Bardeen black hole have remarkable difference from that of the Kerr case.  In Fig.\ref{kerr}, the impact of the rotation parameter $a$ on the $n=1$ scalar QNMs in Kerr, rotating Hayward and Bardeen black holes is shown. We have the same conclusion for the $n=1$ QNMs as the fundamental ones.

\section{QNMs, Lyapunov exponents and shadow radii}

It is known that there is a connection between the QNMs of black holes in the eikonal limit and the unstable null geodesics of the black holes. For the spherical,stationary and asymptotic flat black holes, the real and imaginary parts of the QNMs are multiple of the angular frequency and Lyapunov exponent of the null geodesics, respectively \cite{Ferrari:1984zz,Cardoso:2008bp}. Recently, the extension of the above connection to rotating Kerr black holes \cite{Jusufi:2020dhz,Yang:2012he} and rotating regular black holes \cite{Pedrotti:2024znu}  has also been achieved. In this section, we will check this connection by using our numerical results of the QNMs.

First, let's check the connection between the imaginary parts of the scalar QNMs and the Lyapunov exponents. 
The Lyapunov exponents of circular null geodesics in the equatorial plane of the two black holes are calculated in Section 2. 
The corotating and counterrotating  Lyapunov exponents in rotating Hayward black holes are shown in Table \ref{LEH} and Table \ref{LEHco}, respectively. 
The corotating and counterrotating  Lyapunov exponents in rotating Bardeen black holes are shown in Table \ref{LEB} and Table \ref{LEBco}, respectively. 
The corresponding equatorial QNMs are the ones with $m=\pm l$, where $m=l$ is corotating and $m=-l$ is counterrotating.  
The relation between the Lyapunov exponents and the imaginary parts of QNMs frequency is 
\begin{equation}\label{Im-ly}
	\text{Im}(\omega)=\lambda(n+\frac{1}{2}),\quad n=0,1,2,3\cdots.
\end{equation}
In Table \ref{LEHcom1}, we show the comparison between the corotating Lyapunov exponents and imaginary parts of the fundamental and $n=1$ equatorial QNMs with $m=l=12$. The rotating parameter is chosen as $a=0.2$ and various values of the deviation parameter $g$ are considered to show the impact of $g$ on the relation \eqref{Im-ly}. The QNMs calculated with WKB method and matrix method are denoted as $\omega_{\text{WKB}}$ and $\omega_{\text{MM}}$, respectively. We also show the difference between the multiple of Lyapunov exponents and $-\text{Im}(\omega_{\text{MM}})$, i.e. $\delta=\frac{-\text{Im}(\omega_{\text{MM}})-\lambda(n+1/2)}{\lambda(n+1/2)}$. We can see that the numerical results satisfy the relation \eqref{Im-ly} at one-thousandth level. 
The comparison between the counterrotating Lyapunov exponents and imaginary parts of the fundamental and $n=1$ equatorial QNMs with $m=-l=-12$ is shown in Table \ref{LEHcom2}.    
We can also see that the numerical results satisfy the relation \eqref{Im-ly} quite well, at one-thousandth level. 

In Table \ref{LEBcom1} and Table \ref{LEBcom2}, we show the comparison of Lyapunov exponents and imaginary parts of QNMs in rotating Bardeen black holes for corotating and counterrotating case, respectively. For the corotating case, it is noticed that when the deviation parameter $g_*$ is small enough, the numerical results satisfy the relation \eqref{Im-ly} at one-thousandth level, however, when $g_*$ is around 0.7, the relation \eqref{Im-ly} is satisfied  at several-thousandths level. For the counterrotating case,  
the relation \eqref{Im-ly} is satisfied well at one-thousandth level.

Next, let's consider the connection between the real parts of the QNMs and radii of the black hole shadows in rotating Hayward and Bardeen black holes. This connection for the fundamental QNMs of the two rotating regular black holes was studied in \cite{Pedrotti:2024znu}. Here we just check the connection for the $n=1$ QNMs. On the one hand, the shadow radius $R_{s}$ can be obtained from the real parts of eikonal QNMs by using the follow equation \cite{Pedrotti:2024znu} 
\begin{align}\label{RsfromQNM}
	R_{s}(\mu) &= \sqrt{\left(\frac{1}{\Omega_{R}} - \varepsilon(\mu, a)\right)^{2} - 		\frac{a^{2}}{2}\left(1 - \mu^{2}\right)} \\
	&\approx \sqrt{\left(\frac{\ell + 1/2}{\Re\left(\omega_{n\ell m}\right)} \right)^{2} - 	\frac{a^{2}}{2}\left(1 - \mu^{2}\right)},
\end{align}
where $\mu=\frac{m}{l+1/2}=\frac{L_z}{L}$ and $\Re\left(\omega_{n\ell m}\right)$ represents the real part of QNMs.
On the other hand, the shadow radius $R_{s}$ can also be computed by using closed photon orbits with the following equation
\begin{equation}\label{rrs}
	R_s=\frac{\sqrt{\mathcal{D}+L_z^2}}{E}=\sqrt{\frac{\mathcal{D}}{E^2}+\frac{L^2_z}{E^2}},
\end{equation} 
and together with the Bohr-Sommerfeld quantization condition \cite{Pedrotti:2024znu} 
\begin{equation}\label{bsc}
	2\int_{\theta_-}^{\theta_+}d\theta \sqrt{\Theta} =2\pi(L-|L_z|),
\end{equation}
where $\mathcal{D}$ is Carter's constant, $\Theta=\mathcal{D}-\cos^2\theta\left(\frac{L_z^2}{\sin^2\theta}-a^2E^2\right)$, and $\theta_\pm$ satisfy $\Theta(\theta_\pm)=0$.

\begin{figure}[ht]
	\centering
	\begin{minipage}{0.45\textwidth}
		\includegraphics[width=\linewidth]{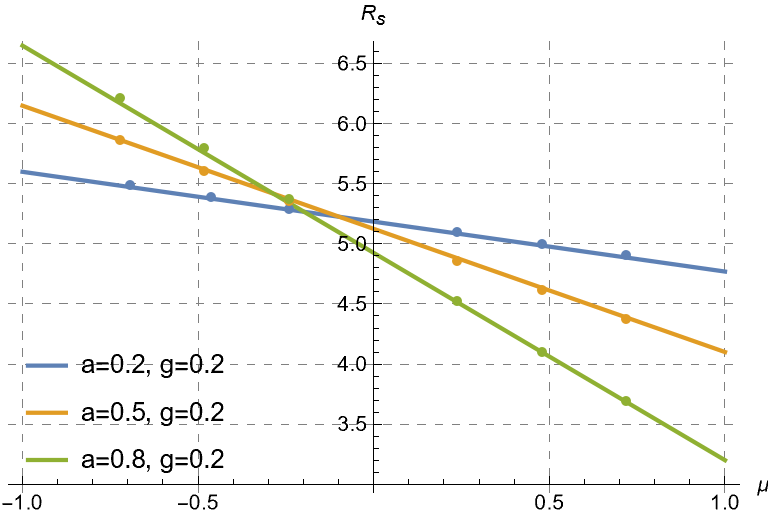}
		
	\end{minipage}\hfill
	\begin{minipage}{0.45\textwidth}
		\includegraphics[width=\linewidth]{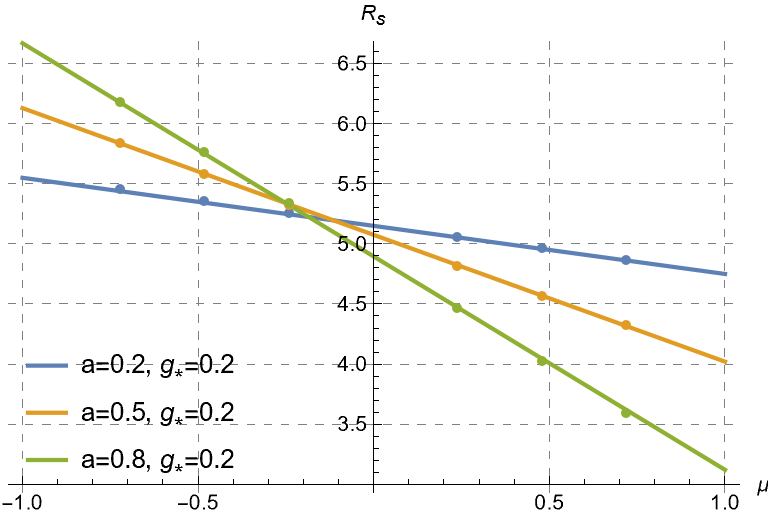}
		
	\end{minipage}
	\caption{The shadow radius directly calculated by using closed photon orbits \cite{Pedrotti:2024znu} are shown in lines. The shadow radius computed from the real parts of $n=1$ QNMs  are show in dots with $l=12,m=\pm3,\pm6,\pm9$. The left is for the rotating Hayward case and the right is for rotating Bardeen case.}
	\label{Rs}
\end{figure}

In Fig.\ref{Rs}, we numerically plot the shadow radius by using equations \eqref{rrs} and \eqref{bsc} with $g=g_*=0.2, l=12$ and various rotation parameters  
for the two rotating regular black holes. The dots in Fig.\ref{Rs} are calculated based the real parts of the $n=1$ QNMs with $l=12, m=\pm3,\pm6,\pm9$ by using equation \eqref{RsfromQNM}. It is noticed that the results of two methods are quite consistent.

\begin{table}[h]
	\centering
	\caption{Comparison between corotating Lyapunov exponents and imaginary parts of QNMs in rotating Hayward black holes. $a=0.2$, $m=l=12$.}
	\label{LEHcom1}
	\begin{tabular}{|c|c|c|c|c|c|c|c|c|}
		\hline
		& \multicolumn{4}{c|}{\textbf{n=0}}   & \multicolumn{4}{c|}{\textbf{n=1}}   \\\hline
		\backslashbox{$g$}{$n$} 
		& $\lambda(n+\frac{1}{2})$ 
		& $-\text{Im}(\omega_{\text{WKB}})$ 
		& $-\text{Im}(\omega_{\text{MM}})$ 
		& $\delta$(\%)  
		& $\lambda(n+\frac{1}{2})$ 
		& $-\text{Im}(\omega_{\text{WKB}})$ 
		& $-\text{Im}(\omega_{\text{MM}})$ 
		& $\delta$(\%) \\\hline
		0.1 & 0.095864 & 0.096038 & 0.095884  & 0.021 & 0.287591 & 0.288292 & 0.287830 & 0.083 \\\hline
		0.2 & 0.095791 & 0.095966 & 0.095812  & 0.022 & 0.287372 & 0.288074 & 0.287616 & 0.085 \\\hline
		0.3 & 0.095590 & 0.095767 & 0.095613 & 0.024 & 0.286769 & 0.287476 & 0.287022 & 0.088 \\\hline
		0.4 & 0.095189 & 0.095370 & 0.095216 & 0.028 & 0.285566 & 0.286282 & 0.285820 & 0.089 \\\hline
		0.5 & 0.094498 & 0.094687 & 0.094533  & 0.037 & 0.283493 & 0.284226 & 0.283769 & 0.097 \\\hline
		0.6 & 0.093391 & 0.093593 & 0.093439  & 0.051 & 0.280172 & 0.280932 & 0.280488 & 0.089 \\\hline
		0.7 & 0.091667 & 0.091890 & 0.091737 & 0.076 & 0.275000 & 0.275809 & 0.275349 & 0.127 \\\hline
	\end{tabular}
\end{table}

	\begin{table}[h]
		\centering
		\caption{Comparison between counterrotating Lyapunov exponents and imaginary parts of QNMs in rotating Hayward black holes. $a=0.2$, $m=-l=-12$.}
		\label{LEHcom2}
		\begin{tabular}{|c|c|c|c|c|c|c|c|c|}
			\hline
			& \multicolumn{4}{c|}{\textbf{n=0}} & \multicolumn{4}{c|}{\textbf{n=1}} \\\hline
			\backslashbox{$g$}{$n$} 
			& $\lambda(n+\frac{1}{2})$ 
			& $-\text{Im}(\omega_{\text{WKB}})$ 
			& $-\text{Im}(\omega_{\text{MM}})$ 
			& $\delta$(\%) 
			& $\lambda(n+\frac{1}{2})$ 
			& $-\text{Im}(\omega_{\text{WKB}})$ 
			& $-\text{Im}(\omega_{\text{MM}})$ 
			& $\delta$(\%) \\\hline
			0.1 & 0.095982 & 0.096138 & 0.096005  & 0.024 & 0.287945 & 0.288675 & 0.288276 & 0.115 \\\hline
			0.2 & 0.095945 & 0.096101 & 0.095968  & 0.024 & 0.287836 & 0.288564 & 0.288164 & 0.114 \\\hline
			0.3 & 0.095846 & 0.096000 & 0.095867 & 0.022 & 0.287538 & 0.288259 & 0.287859 & 0.112 \\\hline
			0.4 & 0.095650 & 0.095800 & 0.095666 & 0.017 & 0.286950 & 0.287656 & 0.287256 & 0.107 \\\hline
			0.5 & 0.095318 & 0.095461 & 0.095328 & 0.011 & 0.285953 & 0.286635 & 0.286236 & 0.099 \\\hline
			0.6 & 0.094800 & 0.094933 & 0.094800  & 0.000 & 0.284400 & 0.285043 & 0.284644 & 0.086 \\\hline
			0.7 & 0.094030 & 0.094147 & 0.094014 & 0.017 & 0.282089 & 0.282673 & 0.282274 & 0.066 \\\hline
		\end{tabular}
	\end{table}

\begin{table}[h]
	\centering
	\caption{Comparison between corotating Lyapunov exponents and imaginary parts of QNMs in rotating Bardeen black holes. $a=0.2$, $m=l=12$.}
	\label{LEBcom1}
	\begin{tabular}{|c|c|c|c|c|c|c|c|c|}
		\hline
		& \multicolumn{4}{c|}{\textbf{n=0}} & \multicolumn{4}{c|}{\textbf{n=1}} \\\hline
		\backslashbox{$g_*$}{$n$} 
		& $\lambda(n+\frac{1}{2})$ 
		& $-\text{Im}(\omega_{\text{WKB}})$ 
		& $-\text{Im}(\omega_{\text{MM}})$ 
		& $\delta$(\%)  
		& $\lambda(n+\frac{1}{2})$ 
		& $-\text{Im}(\omega_{\text{WKB}})$ 
		& $-\text{Im}(\omega_{\text{MM}})$ 
		& $\delta$(\%) \\\hline
		0.1 & 0.095710 & 0.0958863 & 0.095732  & 0.023 & 0.287130 & 0.287836 & 0.287374 & 0.085 \\\hline
		0.2 & 0.095190 & 0.0953734 & 0.095218  & 0.030 & 0.285570 & 0.286291 & 0.285828 & 0.090 \\\hline
		0.3 & 0.094220 & 0.0944169 & 0.094260 & 0.042 & 0.282660 & 0.283411 & 0.282942 & 0.100 \\\hline
		0.4 & 0.092596 & 0.092816 & 0.092656 & 0.065 & 0.277787 & 0.278590 & 0.278113 & 0.117 \\\hline
		0.5 & 0.089872 & 0.091910 & 0.089971  & 0.109 & 0.269615 & 0.270519 & 0.270031 & 0.154 \\\hline
		0.6 & 0.084926 & 0.085278 & 0.085112  & 0.219 & 0.254778 & 0.255909 & 0.255411 & 0.248 \\\hline
		0.7 & 0.073580 & 0.074250 & 0.074051 & 0.639 & 0.220740 & 0.222748 & 0.222044 & 0.591 \\\hline
	\end{tabular}
\end{table}

\begin{table}[h]
	\centering
	\caption{Comparison between counterrotating Lyapunov exponents and imaginary parts of QNMs in rotating Bardeen black holes. $a=0.2$, $m=-l=-12$.}
	\label{LEBcom2}
	\begin{tabular}{|c|c|c|c|c|c|c|c|c|}
		\hline
		& \multicolumn{4}{c|}{\textbf{n=0}} & \multicolumn{4}{c|}{\textbf{n=1}} \\\hline
		\backslashbox{$g_*$}{$n$} 
		& $\lambda(n+\frac{1}{2})$ 
		& $-\text{Im}(\omega_{\text{WKB}})$ 
		& $-\text{Im}(\omega_{\text{MM}})$ 
		& $\delta$(\%)  
		& $\lambda(n+\frac{1}{2})$ 
		& $-\text{Im}(\omega_{\text{WKB}})$ 
		& $-\text{Im}(\omega_{\text{MM}})$ 
		& $\delta$(\%) \\\hline
		0.1 & 0.095912 & 0.096066 & 0.095933  & 0.021 & 0.287735 & 0.288460 & 0.288059 & 0.113 \\\hline
		0.2 & 0.095676 & 0.095826 & 0.095692  & 0.017 & 0.287027 & 0.287734 & 0.287331 & 0.106 \\\hline
		0.3 & 0.095247 & 0.095389 & 0.095253 & 0.007 & 0.285740 & 0.286414 & 0.286007 & 0.094 \\\hline
		0.4 & 0.094559 & 0.094687 & 0.094550 & -0.010 & 0.283677 & 0.284297 & 0.283884 & 0.073 \\\hline
		0.5 & 0.093490 & 0.093595 & 0.093455 & -0.038 & 0.280470 & 0.281003 & 0.280583 & 0.040 \\\hline
		0.6 & 0.091801 & 0.091865 & 0.091722  & -0.086 & 0.275403 & 0.275788 & 0.275360 & -0.016 \\\hline
		0.7 & 0.088967 & 0.088953 & 0.088806 & -0.181 & 0.266901 & 0.267016 & 0.266575 & -0.122 \\\hline
	\end{tabular}
\end{table}

\section{conclusion}
In this work, the fundamental and $n=1$ QNMs of massless scalar field in rotating Hayward and Bardeen regular black holes are studied. 
We calculate the QNMs with two methods: WKB method and matrix method. Since the accuracy of matrix method is higher, most numerical QNM results shown in the Tables and Figures are based on matrix method. We discuss the impacts of the deviation parameters $g$ and $g_*$ on the real and imaginary parts of the QNMs in detail. It is found that compared to the Kerr black hole case, the deviation parameters just lead to percent-level increase for the values of the QNMs. The impact of multipole number $l$ and rotation parameter $a$ on the QNMs are also discussed.     

We also calculate the corotating and counterrotating equatorial Lyapunov exponents of the null circular geodesics in the two rotating regular black holes in detail. 
Then, the connection between the imaginary parts of the QNMs and the equatorial Lyapunov exponents for the two regular black holes are explicitly checked. 
For the fundamental and $n=1$ QNMs, good agreements with equatorial Lyapunov exponents has been found. The difference is at one-thousandth percent level.

The connection between the real parts of the fundamental QNMs and the shadow radius for the two rotating regular black holes was discussed in \cite{Pedrotti:2024znu}. Thus, here we just check the connection for the $n=1$ QNMs in the two rotating regular black holes. A good agreement has been found, which can be expected through the observation that the overtone number has small impact on the real parts of the QNMs.    

Our numerical results indicate that the massless scalar QNMs of the two rotating regular black holes have at most percent-level increase in comparison with the Kerr black hole case.
So it is not easy to distinguish the rotating regular black holes and Kerr black hole from scalar QNMs. It will be interesting to further consider the QNMs of gravitational perturbation of the two regular black holes. 
Recently, signal of phase transition has been unveiled in quasinormal modes of regular AdS black holes \cite{Guo:2024jhg}. It will also be interesting to consider the phase transition of the two rotating regular black holes discussed here from their QNMs.

\section*{Acknowledgments}
The authors would like to thank Zi-Yang Huang for valuable discussions. This work is partially supported by Guangdong Major Project of Basic and Applied Basic Research (No.2020B0301030008).

\bibliographystyle{apsrev}
\bibliography{ref1}

\end{document}